\documentstyle[emulateapj,10pt,apjfonts,psfig]{article}
\def\snr{CTB~80}
\def\psr{B1951+32}

\newcommand\HI{H\,{\sc i}}

\def\kms{km~s$^{-1}$}
\def\etal{et~al. }

\lefthead{MIGLIAZZO ET AL.}
\righthead{PROPER MOTION OF PULSAR \psr}

\begin{document}
\title {PROPER MOTION MEASUREMENTS OF PULSAR~\psr\ IN THE SUPERNOVA REMNANT~\snr}
\author{J. M. Migliazzo\altaffilmark{1}, B. M. Gaensler\altaffilmark{2,1}, 
D. C. Backer\altaffilmark{3}, 
B. W. Stappers\altaffilmark{4}, \\
E. van der Swaluw\altaffilmark{5} and
R. G. Strom\altaffilmark {4,6}} 
\altaffiltext{1}{Center for Space Research, 
Massachusetts Institute of Technology, 70 Vassar Street, Cambridge, MA 02139;
jmig@mit.edu}
\altaffiltext{2} {Harvard-Smithsonian Center for Astrophysics, 
60 Garden Street, Cambridge, MA 02138}
\altaffiltext{3} {Department of Astronomy,
University of California, Berkeley, Berkeley, CA 94720}
\altaffiltext{4} {ASTRON, Postbus 2, 7990 AA, Dwingeloo, The Netherlands}
\altaffiltext{5} {Dublin Institute for Advanced Studies, 
5 Merrion Square, Dublin 2, Ireland}
\altaffiltext{6} {Astronomical Institute `A. Pannekoek,' University of 
Amsterdam, The Netherlands}


\begin{abstract}

Using the Very Large Array and the Pie Town antenna, we have measured
the position of the radio pulsar \psr\ relative to nearby background
radio sources at four epochs between 1989 and 2000. These data show  a
clear motion for the pulsar of $25\pm4$~milliarcsec~yr$^{-1}$ at a
position angle $252^\circ \pm 7^\circ$ (north through east), corresponding
to a transverse velocity $240\pm40$~\kms\ for a distance to the source
of 2~kpc. The measured direction of motion confirms that the pulsar is
moving away from the center of its associated supernova remnant, 
the first time that
such a result has been demonstrated.  Independent
of assumptions made about the pulsar
birth-place, we show that the measured proper motion
implies an age for the pulsar of $64\pm18$~kyr, somewhat less than
its characteristic age of 107~kyr.  This discrepancy can be
explained if the initial spin period of the pulsar was $P_0 = 27\pm6$~ms.

\end{abstract}

\keywords{
ISM: individual (\snr) ---
pulsars: individual (\psr) --- 
radio continuum: ISM ---
stars: neutron ---
supernova remnants
}

\section{Introduction}
\label {sec_intro}

Associations between pulsars and supernova remnants (SNRs) allow
measurements which would not be possible on either population of object
alone. Since the center of an SNR marks the presumed pulsar
birth-site, the pulsar's characteristic age, combined with the offset
of the pulsar from the center of the SNR, lets us estimate the pulsar's
transverse velocity (\cite{fgw94}). A more difficult measurement is
to actually measure such a pulsar's proper motion.  The direction of
motion can confirm (or refute) the association with the SNR, while the
magnitude of the motion gives an independent estimate of the pulsar's
age (\cite{gf00}).

PSR~\psr\ is a rapidly spinning ($P=39.5$~ms) radio,
X-ray, and $\gamma$-ray pulsar, located on the
edge of the unusual SNR~CTB~80 (G69.0+2.7)
(\cite{str87}; \cite{kcb+88}).
While \psr\ is spinning only slightly slower than the Crab pulsar,
its low period derivative implies a much larger characteristic
age, $\tau_c \equiv P/2\dot{P} = 107$ kyr, and
a much lower surface magnetic field,
$B \approx 5\times10^{11}$~G (\cite{ftb+88}).

The distance to the pulsar of $2.4\pm0.2$~kpc, as estimated from its
dispersion measure, is consistent with the SNR's distance of 2~kpc as
measured with \HI\ absorption (\cite{ss00}), suggesting that the two
sources are physically associated. However, this system appears quite
different from more typical pulsar/SNR associations, in which a young
($\tau_c \la 20$~kyr) pulsar sits near the center of an approximately
circular shell.  This difference can  be understood if one invokes an
evolutionary scenario in which the pulsar was originally interior to
the SNR, but as a result of its high space velocity has caught up with
and begun to penetrate its SNR (\cite{fss88}; \cite{hk88};
\cite{krh+90}).  The pulsar's relativistic wind then interacts with the
SNR shell, re-energizing and distorting it (\cite{sfs89}).  
This sequence can explain
both the strange appearance of \snr\ and the large age for
\psr\ compared to most pulsar/SNR associations.

While this evolutionary picture is aesthetically pleasing, it is
imperative that it be verified by observations.  Specifically, the model
proposed for \snr\ can be tested by measurement of the proper motion of
PSR~\psr. If the two sources are associated, then the pulsar should be
moving away from the center of the SNR, as defined by the shell seen
in infra-red and \HI\ (\cite{fss88}; \cite{krh+90}).  The separation
between the pulsar and the shell's center, when combined with the pulsar's
characteristic age, lets us predict a proper motion for the pulsar of
$\mu \approx 15$~mas~yr$^{-1}$ at a position angle (PA) $\approx250^\circ$
(north through east), independent of the distance to the system.

PSR~\psr\ shows significant timing noise, which prevents measurement
of its proper motion through time-of-arrival analysis (\cite{flsb94}).
Thus, only via interferometric measurements can this prediction
be tested. In this {\em Letter}, we report on measurements of the
motion of PSR~\psr\ over an 11-year baseline. In \S\ref{sec_obs} we describe
our observations, while in \S\ref{sec_meas} 
we explain the method used to make our
measurements and present our results. In \S\ref{sec_disc} 
we discuss the implications
of these results for the pulsar and SNR.

\section{Observations}
\label{sec_obs}

Observations of PSR~\psr\ were made with the Very Large Array (VLA) at
epochs 1989.04, 1991.55, 1993.02 and 2000.90.  Each observation was
eight hours in duration and used the VLA's A configuration. In the
1989, 1991, and 1993 observations, data were recorded at two
simultaneous frequencies, 1385 \& 1652 MHz, within the 20-cm band.
Each frequency consisted of 15 channels spread across a 23.5 MHz
bandwidth with a sampling time of 10~seconds, resulting in a useable
field-of-view of radius $\sim10'$ centered on the pulsar.  For the
2000-epoch observations, we also incorporated data from the Pie Town
antenna of the VLBA, which doubled the resolution of the array
primarily in one dimension.  The observations for that epoch were
conducted at 1385 \& 1516 MHz, and consisted of 15 channels across a
12.5 MHz bandwidth with 5-sec sampling, again resulting in a 10-arcmin
field-of-view.  The flux density scale of our observations was
determined using observations of 3C~286, while the time-varying gains
for each antenna were measured using regular observations of
PKS~B1923+210 or TXS~2013+370.

After standard editing and calibration, we produced
images of the pulsar field for each epoch and frequency,
using multi-frequency synthesis to mitigate
bandwidth smearing, and discarding all baselines
shorter than 10~km (corresponding to 
spatial scales larger than $4''$).
By removing these shorter baselines, we ensured that emission 
from SNR~\snr\ and from the compact wind-driven nebula surrounding
the pulsar were not detected.
The only emission
seen in our images was the pulsar itself, the
``hot-spot'' seen immediately adjacent (\cite{str87}), and
various other point sources spread throughout the field.

We have identified seven such sources within $10'$ of the pulsar, as
listed in Table~\ref{tab_pos}.  After deconvolving each image using the
{\tt CLEAN} algorithm, we applied a gaussian fit to the pulsar and to
each of these seven sources to measure their position and extent at each
epoch and frequency. 

\section{Measurement of the Pulsar Proper Motion}
\label{sec_meas}

In order to accurately measure the pulsar proper motion, we adopted six
of the field sources as reference sources, reserving the 
nearest source to the pulsar (source~1 in Table~\ref{tab_pos})
as a check on our measurements. The reference sources
are approximately evenly distributed throughout the field.
The quality of each source as an astrometric reference was determined by
measuring the vector separation between all possible pairs of sources,
and then verifying that no source showed trends or significant changes
in the magnitude and direction of these vectors.  We also  verified that
in all cases the dimensions of the fitted gaussian matched those of the
synthesized beam, confirming that scatter broadening and
bandwidth smearing (in the case of background sources) 
or scintillation (in the case of the pulsar) were not
affecting our positional determinations.

The ionosphere can potentially distort
the measured positions, producing a frequency-dependent
angular displacement
$\Delta\theta=k \theta \lambda^2$, where
$\theta$ is the angular separation between a source
and the field center, $\lambda$ is the observing
wavelength, and $k$ is a constant. By measuring $\theta$
for each source and each frequency, we found 
that $k$ had a value consistent with zero, demonstrating
that any ionospheric effects were dominated by other
uncertanties in  our measurements.

McGary \etal\ (2001\nocite{mbf+01}) 
used the VLA to measure pulsar proper motions
an order of magnitude smaller than that expected here, and found
a variety of other ways in which the positions of reference sources
can be distorted, including relativistic effects due to the Earth's
motion, errors introduced by the VLA correlator and additional empirical
corrections which had no simple explanation.
At the lower precision required here, we have accounted 
for all these effects 
simply by measuring the scatter in the position of each
reference source between epochs.

We began by determining the vector distances between all possible pairings
of the six reference sources. The standard deviations of the components
of each vector in Right Ascension and Declination were computed for the
four epochs; these were taken as the uncertainty for a particular pairing.
This uncertainty was then decomposed between the two sources in that pair,
the relative contributions of the two sources to the joint error 
being weighted
inversely by each source's signal-to-noise ratio.  We were thus able to
derive an uncertainty in the position of each source for each pairing;
these uncertainties were averaged over the five possible pairings
to determine the best measurement of the true uncertainty in each
coordinate for each source.  This analysis was performed separately for
observations at 1385~MHz and at 1516/1652~MHz.  A mean reference position
for each epoch and frequency was then determined by averaging together
the positions of the six reference sources, weighting the contribution
from each source inversely by its distance from the field center.

\begin{figure}[thb]
\vspace{1cm}
\centerline{\psfig{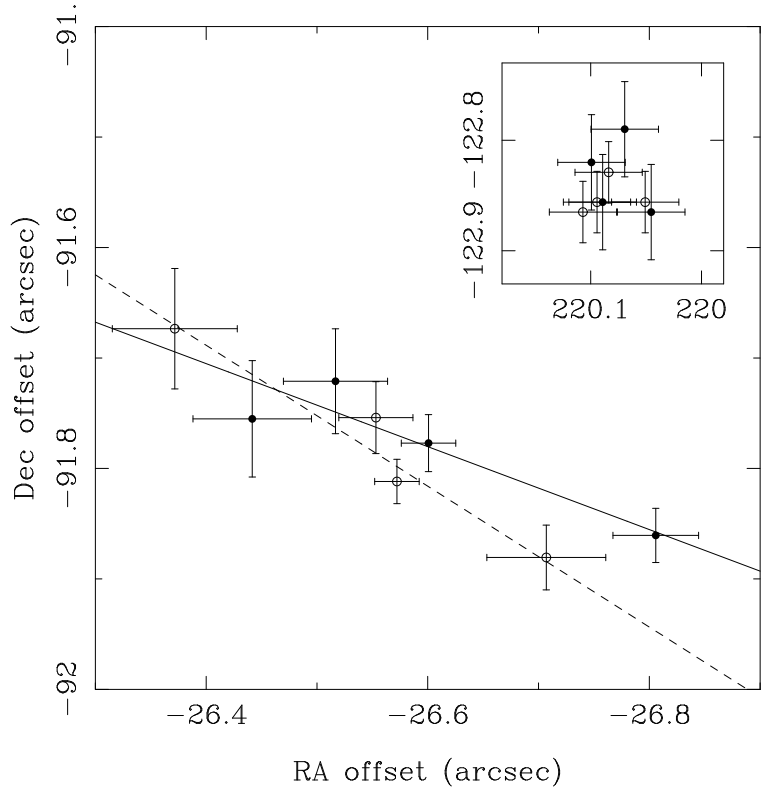}}
\caption{Proper motion measurements of PSR~\psr. 
The data points show the offset of the pulsar from
the mean reference position at four epochs, and the lines show
the weighted best fit to these data.  The solid circles and solid line
correspond to 1385-MHz data, while the open circles and dashed line
represent 1516/1652-MHz data. From left to right, the data points
correspond to observations at epochs 1989.04, 1991.55, 1993.02 and 2000.90.
The inset shows the position of source~1 at each epoch, plotted
on the same scale and determined via the same process.}
\label{fig_motion}
\end{figure}

The separation between the pulsar's position and this mean reference
position was then measured for each epoch and frequency, combining in
quadrature the calculated uncertainties in the pulsar position with
those determined for the reference
position. The results of these measurements are plotted in
Figure~\ref{fig_motion}, and show a clear motion of the pulsar to the
south-west. The flux density of the pulsar was higher during
the 1993 observations than at other epochs, resulting in smaller
uncertainties in the pulsar position for this measurement. Note 
that the position of the pulsar listed in Table~\ref{tab_pos} is
consistent with previous determinations (\cite{flsb94}), but with
larger errors due to the systematic errors discussed above. Such
uncertainties in the absolute astrometry do not affect our proper
motion measurement, which has been determined based on 
astrometry relative to nearby background sources. 

As an independent test of our approach, we have similarly measured the
proper motion for source~1.  As is shown in Figure~\ref{fig_motion}, no
change in position is seen for this source, demonstrating that the
motion measured for the pulsar is real and that the uncertainties have
been realistically assessed.

By applying a weighted least squares fit to the pulsar's position at
each epoch, we find a proper motion at 1385~MHz of  $\mu =
31\pm5$~mas~yr$^{-1}$, at PA~$=250^\circ \pm7^\circ$, and at
1516/1652~MHz of $\mu = 26\pm6$~mas~yr$^{-1}$ at PA~$=240^\circ \pm
10^\circ$.  These two measurements are consistent with each other
within their uncertainties. After taking into account apparent motion
of the pulsar at the level of 5~mas~yr$^{-1}$ due to differential
Galactic rotation,\footnote{To make this calculation, we have used the
rotation curve of Fich, Blitz \& Stark (1989\nocite{fbs89}) and have
assumed a 10\% uncertainty in the systemic velocity of the pulsar.}
the combination of our two measurements yields a motion $\mu =
25\pm4$~mas~yr$^{-1}$ at PA~$=252^\circ \pm 7^\circ$. The corresponding
projected velocity is $V_t = (240\pm40)d_2$~\kms\ for a distance
$2d_2$~kpc.

\section{Discussion}
\label{sec_disc}

\begin{figure}[bht]
\centerline{\psfig{file=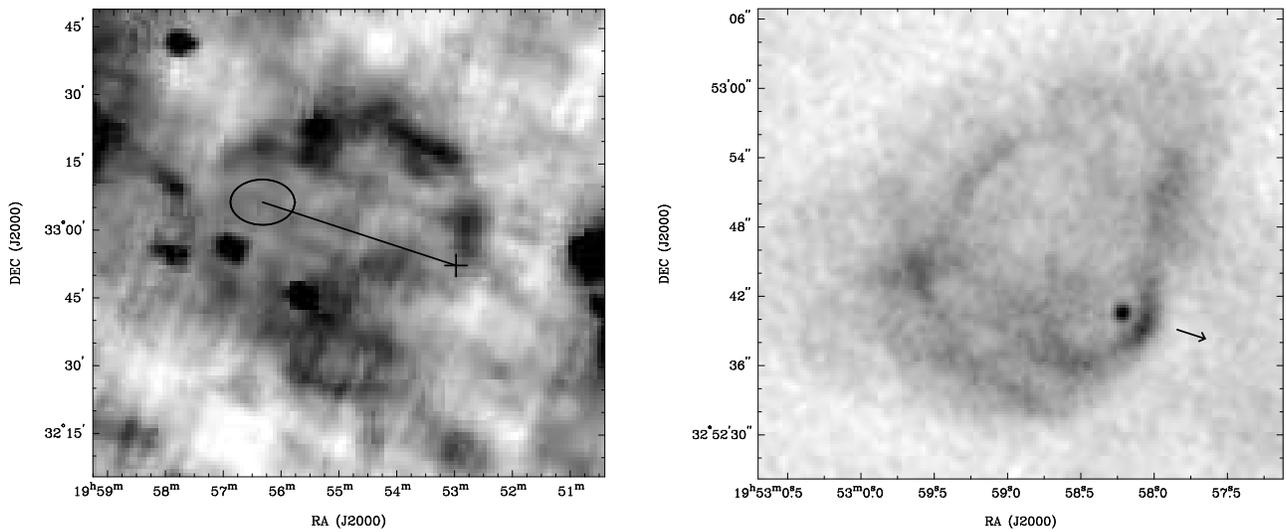,width=17cm}}
\caption{SNR~\snr\ and PSR~\psr. The left panel presents an
image of the {\em IRAS}\ 60/100~$\mu$m ratio in the region, showing
the shell of CTB~80.
The current position of PSR~\psr\ is marked by the ``+'' symbol, the
direction along which it has traveled is indicated by the solid line,
and its projected birthplace for an age of 107~kyr is marked by the
$1\sigma$ ellipse. In the right panel, we show a 1.5-GHz VLA image of the immediate
vicinity of the pulsar, made from all baselines of
our 1993-epoch data. PSR~\psr\ is the
point-source to the south-west of center. The arrow indicates the
pulsar's measured direction of motion, the length of
the arrow corresponding to the distance
traveled by the pulsar over 100~years.} 
\label{fig_ctb80}
\end{figure}

The transverse velocity we have inferred for PSR~\psr\ is in 
agreement with the value $V_t \sim 300$~\kms\ implied by scintillation in
its dynamic spectrum (\cite{ftb+88}).  Furthermore, 
as shown in Figure~\ref{fig_ctb80} the measured position
angle of the proper motion agrees closely with the direction predicted
if the adjacent ``hot-spot'' is interpreted as a bow-shock
driven ahead of the pulsar (\cite{str87}).
It can also be seen in Figure~\ref{fig_ctb80}
that the pulsar is
clearly traveling directly away
from the center of its SNR.  While this result is not unexpected given
the already strong evidence for an association between the pulsar and
SNR, it is surprising to realize that this is the first time such an
effect has actually been observed.  Such measurements for other young
pulsars in SNRs have failed to demonstrate such motion, and have either
required
specific models of SNR and pulsar evolution to maintain the association
(\cite{bmk+89}; \cite{gf00}), or have
caused the association to be abandoned (\cite{tc94}).

While the direction in which the pulsar is moving is as predicted, the
magnitude of the pulsar's motion is not in agreement with
the simplest expectations.  By projecting the pulsar's position back by
$\tau_c=107$~kyr to its inferred birth-site, it can be seen from
Figure~\ref{fig_ctb80} that if the pulsar was born near
the center of its SNR, then the pulsar's true age, $t_p$, must
be less than $\tau_c$.

We can make an initial estimate of $t_p$ by simply identifying the
offset, $\Theta$, of the pulsar from the SNR's center.  We
estimate from the morphology of the infra-red shell shown in
Figure~\ref{fig_ctb80} that  $\Theta \sim 27'$, so that $t_p =
\Theta / \mu \sim 64$~kyr.  However, a high velocity progenitor
and/or asymmetric SNR expansion can produce a significant offset
between the pulsar birth-site and the geometric center of the SNR
(\cite{rtfb93}; \cite{dj96}; \cite{hp99}). Given the complicated
evolutionary history of \snr, it is perhaps overly optimistic to define
a precise geometrical center and simply assume the pulsar was born
there.

If the birth-site is not at the SNR's geometric center, there is no
reason to expect the pulsar to be traveling away from this position
(\cite{gva00}). However, the farther the birth-site from the center of
the SNR, the less likely the pulsar should be moving in this direction
by chance.  We can therefore quantify the uncertainty in the pulsar's
birthplace as follows. We assume that 
for a given angular separation $\phi$ from the
SNR's geometric center, it is equally likely that the pulsar was born
at any position on a circle concentric with the SNR and of radius
$\phi$. Thus, for a particular value of $\phi$, we can compute the
fraction of possible birth-sites which are consistent with the pulsar's
direction of proper motion, PA~$=252^\circ \pm 7^\circ$.  By
considering the full range of values $0 \le \phi \le \Theta$, we
can build up a probability distribution of possible pulsar ages, from
the integral of which we can determine the $1\sigma$ confidence limits
on $\phi$. Using this approach, we find that the pulsar was born within
$\pm8'$ of the SNR's center, and that the age implied by our proper
motion measurement is $t_p = 64\pm18$~kyr. These estimates are
independent of the distance to the system and on the site of the
supernova explosion; they depend only on the assumptions that an
approximate geometric center can be defined for the SNR, and that the
pulsar is moving away from its birthplace.  The age we have derived is
consisent with the estimate of 77~kyr, derived from the expansion
velocity of the \HI\ shell coincident with the SNR (\cite{krh+90}).
 
It is not surprising that this system's true age
is less than the pulsar's characteristic age. For a pulsar
of constant moment of inertia and magnetic dipole moment,
it is straightforward to show that (\cite{mt77}):
\begin{equation}
t_p =  \frac{2\tau_c}{n-1} \left[ 1 - \left( \frac{P_0}{P} \right)^{n-1}\right],
\label{eqn}
\end{equation}
where $n$ is the pulsar's ``braking index'' and $P_0$ is its
initial spin-period. For $n=3$ and $P_0 \ll P$, Equation~(\ref{eqn}) 
reduces to $t_p = \tau_c$, but in general neither such condition
will be satisfied. While $n$ has not been measured for
PSR~\psr, we find $1.5 \la n \la 3$ for the five pulsars
which have had their braking indices measured (\cite{zmg+01} and
references therein).
For $n=3$, the range in $t_p$ determined above
implies $P_0 = 25\pm5$~ms for PSR~\psr, while for
$n=1.5$ we find $P_0 = 29\pm3$~ms. Combining these
estimates, we find the initial period most likely
falls in the range $P_0 = 27\pm6$~ms.
Again, this estimate
does not depend on the assumed distance to the source.

As listed in
Table~\ref{psr_period}, 
PSR~\psr\ joins a rapidly growing list of pulsars whose initial periods
have been determined from their associations with either
SNRs or historical supernovae.
While these measurements span a reasonable range of
initial periods, the data provide good evidence
that radio pulsars are born as rapid rotators, with no evidence for
a population of longer initial periods as has been sometimes proposed
(\cite{vn81}; \cite{ce86}). On the other hand, it is
becoming abundantly clear that $\tau_c$ can be an unreliable
estimator of a young pulsar's age, with corresponding implications
for associations with SNRs,  pulsar velocities, and models
for neutron star cooling. With
many new pulsar/SNR associations now being identified at both radio
and X-ray wavelengths,
a statistically
useful sample of pulsar initial spin periods should soon emerge.


\begin {acknowledgements}

We are grateful to  Roger Foster for initiating this project, and to
Dale Frail for making us aware of these archival data.
We thank Robin McGary and Glen Monnelly for useful advice.
The National Radio Astronomy Observatory is a facility of the
National Science Foundation, operated under cooperative agreement by 
Associated Universities, Inc.  J. M. M. acknowledges the support of
MIT through a Presidential Fellowship for graduate studies in Physics.
B. M. G. acknowledges the support of a Clay Fellowship awarded by the
Harvard-Smithsonian Center for Astrophysics. E. vdS. is currently 
supported by the European Commission under the TMR
programme, contract number ERB-FMRX-CT98-0168.

\end{acknowledgements}


\begin{thebibliography}{}

\bibitem[Bailes \etal  1989]{bmk+89}
Bailes, M., Manchester, R.~N., Kesteven, M.~J., Norris, R.~P., \& Reynolds,
  J.~E. 1989, { ApJ}, {\rm 343}, L53.

\bibitem[Camilo \etal  2002]{cmg+02}
Camilo, F., Manchester, R.~N., Gaensler, B.~M., Lorimer, D.~L., \& Sarkissian,
  J. 2002, { ApJ}, {\rm }.
\newblock in press (astro-ph/0201384).

\bibitem[Chevalier \& Emmering 1986]{ce86}
Chevalier, R.~A. \& Emmering, R.~T. 1986, { ApJ}, {\rm 304}, 140.

\bibitem[Dohm-Palmer \& Jones 1996]{dj96}
Dohm-Palmer, R.~C. \& Jones, T.~W. 1996, { ApJ}, {\rm 471}, 279.

\bibitem[Fesen, Shull, \& Saken 1988]{fss88}
Fesen, R.~A., Shull, J.~M., \& Saken, J.~M. 1988, { Nature}, {\rm 334}, 229.

\bibitem[Fich, Blitz, \& Stark 1989]{fbs89}
Fich, M., Blitz, L., \& Stark, A.~A. 1989, { ApJ}, {\rm 342}, 272.

\bibitem[Foster \etal  1994]{flsb94}
Foster, R.~S., Lyne, A.~G., Shemar, S.~L., \& Backer, D.~C. 1994, { AJ}, {\rm
  108}, 175.

\bibitem[Frail, Goss, \& Whiteoak 1994]{fgw94}
Frail, D.~A., Goss, W.~M., \& Whiteoak, J. B.~Z. 1994, { ApJ}, {\rm 437}, 781.

\bibitem[Fruchter \etal  1988]{ftb+88}
Fruchter, A.~S., Taylor, J.~H., Backer, D.~C., Clifton, T.~R., \& Wolszczan, A.
  1988, { Nature}, {\rm 331}, 53.

\bibitem[Gaensler \& Frail 2000]{gf00}
Gaensler, B.~M. \& Frail, D.~A. 2000, { Nature}, {\rm 406}, 158.

\bibitem[Gvaramadze 2000]{gva00}
Gvaramadze, V.~V. 2000, {\rm }.
\newblock astro-ph/0005572.

\bibitem[Hester \& Kulkarni 1988]{hk88}
Hester, J.~J. \& Kulkarni, S.~R. 1988, { ApJ}, {\rm 331}, L121.

\bibitem[Hnatyk \& Petruk 1999]{hp99}
Hnatyk, B. \& Petruk, O. 1999, { A\&A}, {\rm 344}, 295.

\bibitem[Kaspi \etal  2001]{krv+01}
Kaspi, V.~M., Roberts, M. S.~E., Vasisht, G., Gotthelf, E.~V., Pivovaroff, M.,
  \& Kawai, N. 2001, { ApJ}, {\rm 560}, 371.

\bibitem[{Koo} \etal  1990]{krh+90}
{Koo}, B.-C., {Reach}, W.~T., {Heiles}, C., {Fesen}, R.~A., \& {Shull}, J.~M.
  1990, { ApJ}, {\rm 364}, 178.

\bibitem[Kulkarni \etal  1988]{kcb+88}
Kulkarni, S.~R., Clifton, T.~R., Backer, D.~C., Foster, R.~S., Fruchter, A.~S.,
  \& Taylor, J.~H. 1988, { Nature}, {\rm 331}, 50.

\bibitem[Manchester \& Taylor 1977]{mt77}
Manchester, R.~N. \& Taylor, J.~H. 1977, { Pulsars}, (San Francisco: Freeman).

\bibitem[Marshall \etal  1998]{mgz+98}
Marshall, F.~E., Gotthelf, E.~V., Zhang, W., Middleditch, J., \& Wang, Q.~D.
  1998, { ApJ}, {\rm 499}, L179.

\bibitem[{McGary} \etal  2001]{mbf+01}
{McGary}, R.~S., {Brisken}, W.~F., {Fruchter}, A.~S., {Goss}, W.~M., \&
  {Thorsett}, S.~E. 2001, { AJ}, {\rm 121}, 1192.

\bibitem[Murray \etal  2002]{mss+01}
Murray, S.~S., Slane, P.~O., Seward, F.~D., Ransom, S.~M., \& Gaensler, B.~M.
  2002, { ApJ}, {\rm }.
\newblock in press (astro-ph/0108489).

\bibitem[Reynolds 1985]{rey85}
Reynolds, S.~P. 1985, { ApJ}, {\rm 291}, 152.

\bibitem[R\'{o}\.{z}yczka \etal  1993]{rtfb93}
R\'{o}\.{z}yczka, M., Tenorio-Tagle, G., Franco, J., \& Bodenheimer, P. 1993, {
  MNRAS}, {\rm 261}, 674.

\bibitem[Shull, Fesen, \& Saken 1989]{sfs89}
Shull, J.~M., Fesen, R.~A., \& Saken, J.~M. 1989, { ApJ}, {\rm 346}, 860.

\bibitem[Strom 1987]{str87}
Strom, R.~G. 1987, { ApJ}, {\rm 319}, L103.

\bibitem[Strom \& Stappers 2000]{ss00}
Strom, R.~G. \& Stappers, B.~W. 2000, in { Pulsar Astronomy --- 2000 and
  Beyond, {IAU} Colloquium 177}, ed.\ M. Kramer, N. Wex, \& R. Wielebinski,
  (San Francisco: Astronomical Society of the Pacific), p.~509.

\bibitem[Thompson \& C\'ordova 1994]{tc94}
Thompson, R.~J. \& C\'ordova, F.~A. 1994, { ApJ}, {\rm 421}, L13.

\bibitem[Torii \etal  1999]{ttd+99}
Torii, K., Tsunemi, H., Dotani, T., Mitsuda, K., Kawai, N., Kinugasa, K.,
  Saito, Y., \& Shibata, S. 1999, { ApJ}, {\rm 523}, 69.

\bibitem[Vivekanand \& Narayan 1981]{vn81}
Vivekanand, M. \& Narayan, R. 1981, { J. Astrophys. Astr.}, {\rm 2}, 315.

\bibitem[Zhang \etal  2001]{zmg+01}
Zhang, W., Marshall, F.~E., Gotthelf, E.~V., Middleditch, J., \& Wang, Q.~D.
  2001, { ApJ}, {\rm 554}, L177.

\end{thebibliography}

\clearpage
\vspace{1cm}
\begin{table}[hbt]
\caption{Positions of sources at epoch 2000.90. Values
in parentheses indicate the uncertainty in the last digit.}
\label{tab_pos}
\begin{tabular}{cllc} \hline \hline
Source	& \multicolumn{2}{c}{Position (J2000)} & Distance from \\
       & Right Ascension   & Declination & pulsar (arcmin) \\ \hline
PSR~\psr\ & $19^{\rm h}52^{\rm m}58\fs2(1)$  & $+32^\circ52'41''(1)$
& $\ldots$ \\ 
1	& $19^{\rm h}53^{\rm m}17\fs80(1)$   & $+32^\circ52'09\farcs6(1)$ & 4.1 \\
2	& $19^{\rm h}53^{\rm m}16\fs36(1)$   & $+32^\circ48'46\farcs4(1)$ & 5.5 \\
3	& $19^{\rm h}53^{\rm m}15\fs63(1)$   & $+32^\circ59'38\farcs1(1)$ & 7.9 \\
4	& $19^{\rm h}53^{\rm m}25\fs41(1)$   & $+32^\circ58'12\farcs5(1)$ & 8.0 \\
5	& $19^{\rm h}53^{\rm m}13\fs41(1)$   & $+33^\circ01'21\farcs8(1)$ & 9.3 \\
6	& $19^{\rm h}52^{\rm m}15\fs79(1)$   & $+32^\circ49'35\farcs7(1)$	& 9.4 \\ 
7	& $19^{\rm h}52^{\rm m}15\fs64(1)$   & $+32^\circ49'39\farcs2(1)$	& 9.4 \\ \hline \hline
\end{tabular}
\label{tab_obs}
\end{table}

\clearpage

\vspace{1cm}
\begin{table}[hbt]
\caption{Initial period estimates for young pulsars.}
\label{psr_period}
\begin{tabular}{llcl} \hline \hline
Pulsar		& Associated SNR &
Initial Period & Reference	\\ 
      & or Nebula & (ms) & \\ \hline
J0537--6910 & N~157B & $< 14$	& Marshall \etal\ (1998\nocite{mgz+98})	\\
B0531+21 & Crab Nebula & 19 & Manchester \& Taylor (1977\nocite{mt77})	\\
\psr\ 	& CTB~80 & $27\pm6$		& This paper 	\\ 
B0540--69 	& 0540--69.3  & $30\pm8$ & Reynolds (1985\nocite{rey85}) \\
J0205+6449  & 3C~58 & $60$ & Murray \etal\ (2002\nocite{mss+01})\\  
J1811--1925 & G11.2--0.3 & $62$ & 
Torii \etal\ (1999\nocite{ttd+99}); Kaspi \etal\ (2001\nocite{krv+01}) \\
J1124--5916 & G292.0+1.8 & $90$	& Camilo \etal\ (2002\nocite{cmg+02}) \\ 
\hline \hline
\end{tabular}
\end{table}

\clearpage

\vspace{1.5cm}

\end{document}